\documentclass{elsart} 

 \usepackage{graphicx}

\usepackage{amssymb}
\newcommand{\be}{\begin{equation}}
\newcommand{\ee}{\end{equation}}

\begin{document}

\begin{frontmatter}



\title{Where has all the information gone?}


\author{H. D. Zeh}

\address{Universit\"at Heidelberg\\
www.zeh-hd.de}

\begin{abstract}
The existence of spacetime singularities is irrelevant for the
irreversible appearance of black holes. However, confirmation of the
latter's unitary dynamics would require the preparation of a coherent
superposition of a tremendous number of appropriate ``Everett worlds''.
\end{abstract}

\end{frontmatter}

\section {The information loss ``paradox''}
Stephen Hawking's claim of a lost bet~\cite{Hawking} recently stirred up a
lot of  interest and discussion in the media (physical journals included).
If correct, it would mean that the information absorbed by a black hole
must later be emitted in some way, for example by means of correlations
existing within the Hawking radiation -- even though they
can hardly ever be {\it used} to recover the original information. The
opposite assumption that this information is irretrievably lost, while the
black hole  may completely disappear, is generally
regarded as a paradox, since it would violate
unitarity~\cite{Hawk76,tHooft,Preskill,Hor+M}. 
 
Hawking's new arguments against his own bet rely on a
detailed (though not yet published) calculation, which must use certain
assumptions and approximations -- they are {\it not} based on any novel
empirical evidence. Therefore, his result must simply reflect these
assumptions, regardless of whether or not they are explicitly stated. For
example, if unitarity is presumed for the underlying theory, Hawking's claim
does not need any further calculation~\cite{Page}. Similarly, disappearance
of ``information'' would be unavoidable if a classical spacetime that
contained future singularities were assumed to correctly describe the real
world.  

However, exact classical spacetimes (or gravitational fields) are known to
be inconsistent with the presence of quantized
matter~\cite{BohrEin,PageGeilker}. According to canonical quantum gravity
they are even excluded in the same way as particle trajectories are in
quantum mechanics.  Opposite conclusions about the (ir)reversibility of
black hole dynamics thus arise from different {\it beliefs} regarding the
universal validity of quantum theory. 

Albert Einstein's general relativity requires the existence of
spacetime horizons. They lead to the remarkable consequence that quantum
entanglement, which again Einstein brought into particular focus in his
paper with Podolsky and Rosen
\cite{EPR} (albeit in order to prove quantum theory incomplete), must
arise between the inner and outer regions of a black hole. This leaves
the corresponding ``information'' drastically nonlocal, that is, neither
inside nor outside. John Bell's analysis and subsequent experiments have
demonstrated that entanglement must be part of reality rather than
being the consequence of an incomplete description (mere {\it statistical}
correlations). 

The dispute about the nature of an information loss (or entropy increase) is
not at all new. The conflict between deterministic laws and irreversible
phenomena occurred on many occasions. Let me therefore emphasize that
information is here usually defined by means of a formal ensemble of possible
states which are assumed to evolve in time according to some dynamical law.
For example, information would be conserved under
deterministic (or unitary in the case of quantum theory) laws. The problem
then regards such (always hypothetical, though in most cases empirically
verified) laws -- not what {\it we} happen to know, or are able to use,
observe or calculate.

\section {Information loss in classical statistical physics}
Many deviations from information-conserving laws are 
meaningful and successfully used. For example in classical physics, all
master equations, such as Boltzmann's collision equation, are based on the
permanent neglect of arising correlations (or of other kinds of ``irrelevant
information'') when calculating into the future direction of time. The
applicability of this very restrictive assumption, which
requires a special {\it initial} condition for our world, is
responsible for the observed increase of phenomenological entropy. By
definition, {\it ensemble entropy} is conserved under deterministic
equations of motion if it is calculated -- in contrast to phenomenological
entropy -- by taking into account all irrelevant information, such as
correlations. This would require a highly non-extensive concept of entropy,  
\be
S = -k_B \int \rho (p,q) \ln \rho (p,q) \ dpdq \quad , 
\ee
which is a functional of the density $\rho (p,q)$ in many-particle (or
any other canonical) phase space. Statistical correlations between
local objects at different positions are non-local, that is, they are
themselves {\it at no place}. Only if
$\rho$ approximately factorizes into a mean spatial density
and the rest does one obtain a spatial entropy {\it density}, which allows
entropy to flow in space (as phenomenologically required)~\cite{TD}. The
observed entropy increase reflects the transformation of part of the
conserved information into information about irrelevant properties, such as
uncontrollable correlations. The latter are usually neglected by
the concept of physical (local) entropy. However, the {\it real} (completely
defined) physical state does {\it not} contain any statistical
correlations.

Correlations propagate and multiply very efficiently by
interactions with the environment -- either by means of chains of Boltzmann
type collisions between molecules, or by (even very weak) long range
interactions, since the latters' effect on the molecules of a distant gas is
usually strongly amplified by subsequent molecular collisions~\cite{Borel}.
Hence, in a ``causal'' world, where all correlations require {\it local}
causes in their past, the information they may represent is irreversibly
lost for all practical purposes. The deterministic evolution of most
subsystems of the universe depends crucially on the (otherwise irrelevant
and therefore unknown) precise physical states of their global environments
-- even though determinism is classically well  defined in principle by the
presumed global dynamics. An exact reversal of motion (or a ``complete
recovery of information'') for any macroscopic system would therefore
require a precise reversal or recurrence of the state of the whole universe.

\section{``Information'' in quantum theory}

In quantum theory, global unitarity would similarly warrant the
conservation  of global entropy or lacking ``information" (negentropy) if
this is now defined by the functional
\be
S = -(I-I_0) = - k_B {\rm Trace} ( \rho \ln \rho )  
\ee
of a global density matrix $\rho$. However, quantum theory is usually
understood as an indeterministic (probabilistic) theory. It is {\it this}
contrast which forms the most fundamental information loss paradox in
physics.

If one now assumed (with Bohr) that
quantum concepts were not applicable to macroscopic objects, unitarity
would not even be an issue for them. If one assumes instead (with von
Neumann, Pearle or Ghirardi) that quantum concepts are universal, while the
Schr\"odinger equation has to be modified in order to describe the
collapse of the wave function, unitarity is an approximation -- valid only
for microscopic objects. However, if the Schr\"odinger
equation is assumed to be exact, one has to conclude that there is a
superposition of myriads of branching Everett ``worlds". The time arrow
of this branching must again be the consequence of a cosmic initial
condition (for the universal wave function). Our observed universe would then
be represented by one single, dynamically autonomous branch of the global
wave function that permanently splits further by means of the
dislocalization of quantum phases (decoherence
\cite{decoh}). Considered by itself, this branch describes the same
stochastic phenomena as a genuine collapse (that is, {\it quantum
measurements} in a general sense). According to this description, part of
the information about the initial global state is deterministically
transformed into inaccessible quantum correlations between different Everett
branches, while the growing specification of the observer's branch
(characterized by all previous ``measurement results'') describes in
principle a (usually negligible) entropy {\it decrease}. So the crucial
question is: What precisely is the global density matrix used in the above
definition of entropy? In practice, it is always based on a coarse graining
of locality as well as on an effective collapse (Everett branching).

Quantum correlations (entanglement) are very different from
classical statistical correlations. This becomes most obvious when one
attempts to trace back in time the quantum state of a system by using the
new information gained in a measurement (``postselection''). While
in classical physics this procedure would improve knowledge (reduce an
ensemble) also about the past, in quantum theory the postselected state,
if calculated backwards in time, is in general incompatible with the
documented history. The reason is that a quantum density matrix does {\it
not} simply represent an ensemble (a probability distribution) of possible
states -- that is, not just incomplete information. Quantum correlations
characterize superpositions, which depend on the phases of their complex
coefficients. They define {\it individual} physical states. For
example, two particles with definite total spin are in a
superposition of product states by means of Clebsch-Gordon
coefficients (even when at very different locations). A collapse of this
wave function into a product of states would violate angular momentum
conservation. The term ``quantum information'', though intuitively
appealing because of certain analogies with the classical case, should
therefore be used with great care in order to avoid misconceptions.
Nonetheless, entanglement {\it is} the essential concept underlying the
information loss paradox of black holes.

This ``loss of quantum information'' occurs in particular in the form of
decoher\-ence. Thereby, quantum correlations with an
uncontrollable environment are produced in an irreversible manner (in
formal analogy to Boltzmann correlations in classical mechanics). Their
phase relations are then irrelevant for all local observers, who, for the
same reason, would not possess any physical state by their own any more. 
This gives rise to an
 {\it effective ensemble} of local ``states of being conscious'' for each of
them, existing in different Everett ``worlds''. The various states of
different observers are thereby {\it correlated} by means
of their specific entanglement. 

In this way, outcomes of measurements or other consequences of
decoherence are {\it objectivized}, and may as well be described by
stochastic dynamics (effective master or quantum Langevin equations). Without
any explicit dynamical description of measurements, there would be ``no
measurement problem, as there is no theory'' \cite{Z70}. The identification
of a specific result by the subjective observer may then be regarded as its
selection from the apparent ensemble that was {\it created} by decoherence.
It does {\it not} represent the selection of a subensemble from a
pre-existing ensemble (a mere ``increase of information''). For this reason,
the ensemble entropy of the global system may be smaller than the formal
entropy of any of its local subsystems even in the case of complete global
information (that is, a pure global state) -- see Chap.~2 of~\cite{decoh}.

In scattering experiments, the unitary quantum description is usually
interrupted at the detector. This comes very close to the Copenhagen
interpretation, although it can be justified by the unavoidable (unitary)
interaction of the macroscopic detector with its environment. Even the
macroscopic parts of this undoubtedly real object rely in an essential way
on quantum theory. The pragmatic attitude may work for practical purposes --
but it is clearly not conceptually consistent. It is this conceptual defect
that is usually attributed to an ``absence of microscopic reality'' (or of a
``quantum world''). In particular, detectors and other macroscopic objects,
though strongly entangled with their environments, may fall into black holes!

\section{Information Loss in General Relativity}

In general relativity, matter may fall onto singularities within
finite proper times, whereby all information about its state would become
meaningless (or disappear). On the other hand, simultaneities, on which
the global states are defined, may now be {\it chosen} in such a way that
they never reach the local singularity for coordinate values corresponding to
all finite times in the asymptotic region
(see Fig.~1). This choice should {\it not} affect the physics and its correct
description in regions of space where differently chosen simultaneities
coincide.

\begin{figure}[t]
\centering\includegraphics[width=.8\textwidth]{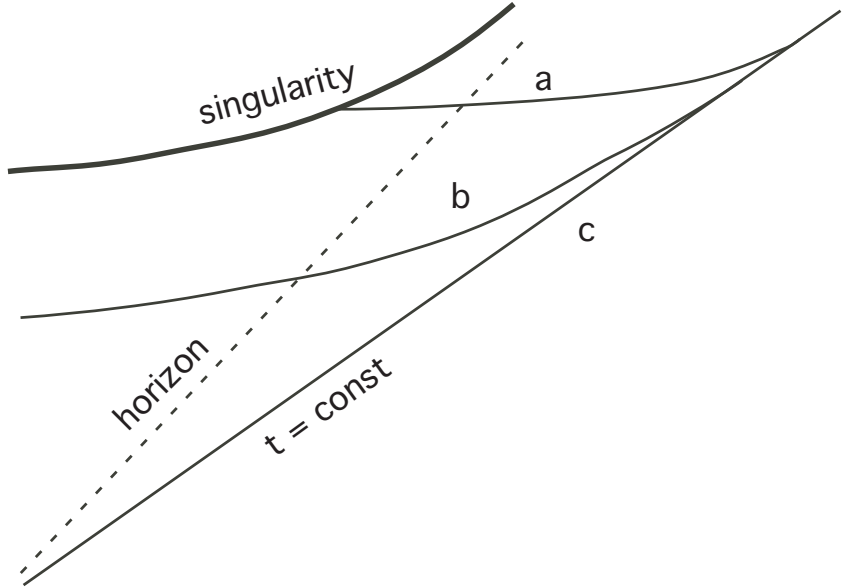}
\caption{{\footnotesize Various simultaneities for a black hole in a
Kruskal type diagram: (a) hitting the singularity, (b) entering the regular
interior region only, (c) completely remaining outside (Schwarzschild time
coordinate $t$). Schwarzschild time is appropriate in particular for
posing asymptotic boundary conditions. The angle between the horizon and
the line $t= const$ can here be arbitrarily changed by a passive time
translation. This includes the (apparently close) vicinity of the horizon,
which can thus be arbitrarily ``blown up'' in the diagram -- thus
focussing consideration on the distant future.}}
\end{figure}

The latter expectation would not represent a conceptual problem if physics
were local. The {\it real} (completely defined) external state of a black
hole, say, would then exist independently of the internal
one, while correlations could affect only a statistical (incomplete)
description. 

In nonlocal quantum theory, however, not only the state of matter on the
singularity, but also its entanglement with that in the regular region is
lost. Although it may be tempting to regard the singularity as an
``ultimate environment'' (which might {\it explain} a genuine collapse of the
wave function), the free choice of simultaneities demonstrates that this
is by no means conclusive. Even the global vacuum
contains entanglement between both sides of a horizon,  locally
giving rise to {\it thermal} Hawking or Unruh radiation (that is, with or
without a singularity).  Since the reduced density matrix that describes
local systems at some distance from the horizon cannot depend
on the continuation of the simultaneity inside the horizon or in its close
external vicinity, it suffices that the ``information'' corresponding to
these quantum correlations is lost for all practical purposes of distant
observers. In principle, there is then no difference to conventional
decoherence  (without black holes)~\cite{KieferCQG2001}. Because
of the extreme time dilation, even the close {\it external} vicinity of the
horizon can causally affect distant systems, such as external observers,
only in the very distant future.

Since the value of Schwarzschild time $t$ (the {\it arcus tangens} of the
angle at which it appears in the figure) has no absolute meaning, the
external part of the diagram remains valid far into the future -- including
times when the black hole should have disappeared by means of Hawking
radiation from the point of view of an external observer. He would observe
the black hole disappear even before (according to these Schwarzschild
simultaneities) a horizon has formed. The mixed state describing the
radiation (a global ensemble) is then created only by the usual
statistical treatment of the radiation process (cf.~\cite{Page}). 

Infalling matter would in turn be affected by cosmic events or
boundary conditions in the distant future if the thermodynamical arrow of
time (``causality'') ever reversed its direction during the cosmic
evolution. Because of the diverging time  dilation this would allow the
matter just in time to feel advanced radiation and experience recoherence
in order to retro-cause the black hole to grow hair and expand again.  For
example, a time-symmetric version of Penrose's Weyl tensor condition
\cite{Penrose} (a vanishing Weyl tensor on {\it all} singularities) would
entirely eliminate all inhomogeneous singularities (such as those of black
holes) together with their event horizons, without essentially affecting the
state of our present universe far from black holes~\cite{Mazagon,KZ}. Global
``information'' (taking into account all correlations) could then be
conserved under a Schr\"odinger equation and for {\it all} choices of
simultaneities. 

It seems that our conventional interpretation of black holes owes too much
to the classical picture that ``nothing unusual happens'' at the
horizon. While this would remain true under a purely local consideration,
observers orbiting and approaching a galactic black hole, for example,
might have sufficient time to observe (and get affected by) much of the
future cosmic history as in an extreme quick motion movie -- provided
they can survive the novel affects caused on them by the external world. 
However, double-ended boundary conditions, such as a symmetric Weyl
tensor condition, may not be dynamically consistent~\cite{antiSchulman} with
classical general relativity, which allows free initial {\it or} final
conditions only.

\section{Quantum Gravity}

The information loss paradox is usually discussed in connection with
quantum gravity. Otherwise the question of unitarity would not make
sense for black holes as geometrodynamical objects. However, regarding
black holes as isolated quantum systems obeying a Schr\"odinger equation
would repeat the popular mistake of describing a quantum measurement
according to von Neumann as an isolated unitary interaction between the
measured system and a macroscopic device. In both cases, decoherence by the
environment is a crucial part of the story.

Just as von Neumann's measurement interaction does not depend on any details
of the apparatus, the key argument regarding black hole unitarity does not
depend on the precise variables which describe quantum gravity -- provided
only the basic principles of quantum theory are maintained. The very idea of
``quantization'' can generally be understood as the conceptual reversal of
decoherence: the re-introduction of those superpositions which are generally
suppressed by interaction with the environment, and which were therefore
missing in the classical description. In general relativity, quantization
thus leads to a wave functional on superspace (the configuration space of
all spatial geometries, which form the kinematics of general relativity). It
is quite irrelevant for the present discussion whether this configuration
space may later have to be modified (or reconstructed) under more general
mathematical considerations -- for example in terms of a
configuration space consisting of loop integrals or strings and other fields
on some higher dimensional space instead of the traditional superspace
including matter fields. All one needs is to take the wave function
seriously~\cite{ZWheeler}. There is neither a ``crisis'' nor an indication
of a ``paradigm shift''~\cite{Susskind} that would not yet have occurred in
a universally valid quantum mechanics. As usual, the paradox is an artifact
of the insistence on classical concepts.

In particular, superpositions of different spatial geometries are permanently
decohered by (become entangled with) matter. In this way, quasi-classical
spacetimes {\it emerge} in the form of propagating, dynamically autonomous
wave packets from the universal wave function~\cite{Joos}. Black holes even
owe their time-{\it asym\-metric} properties, including future horizons, to
this embedding into their time-directed
environment~\cite{KieferCQG2001,Demers}. Since general relativity is time
reversal-symmetric, energy eigenstates would have to be symmetric or
antisymmetric {\it superpositions} of black and ``white'' holes~\cite{HajK},
but would immediately decohere into their components in an irreversible
manner if they ever came into existence. This establishes a superselection
rule separating black and white holes, which, because of the long range of
gravitational interactions, must even be correlated with the time arrow of
the universe (analogous to correlations relating different observers of the
same measurement). Therefore, only {\it black} holes can be {\it observed}.
Interference experiments with black holes, just as with other macroscopic
objects, would require the coherent preparation of many Everett branches
(leading to their recoherence), while microscopic ``virtual holes'' would
neither be black nor white. However, the structure of this Everett branching
may appear very different to asymptotic observers and those being close to
black hole horizons~\cite{inYourgrau}.

The deepest consequence of quantum gravity (and other
repara\-metrization invariant theories) is the absence of any dynamical
time parameter on a fundamental level. This requires that global quantum
states must obey the Wheeler-DeWitt equation $H \Psi = 0$. A
concept of time describing a {\it succession} of states that, in particular,
allows the formal distinction between initial and final conditions can then
only be derived within the range of validity of a Born-Oppenheimer
approximation with respect to the Planck mass~\cite{Zeh86,Kiefer}.
Fundamental cosmic boundary conditions have instead to be postulated (or
derived from new principles) for the timeless state $\Psi$~\cite{CZ}. 

Unless otherwise enforced by means of such boundary conditions,
this stationary solution $\Psi$ of a real Wheeler-DeWitt equation must be
real, too, and may in some regions of configuration space contain a
factor
$\sin(ka)$, characterizing the cosmic expansion parameter $a$. This is then
usually decohered by its environment into components
$e^{ika}$ and
$e^{-ika}$, where the sign of $i$ has no physical meaning in the absence
of any time dependence of the form $\exp(i\omega t)$. Therefore, there is
no distinction between big bang and big crunch any more in
quantum gravity. The Born-Oppenheimer approximation would
break down close to a conceivable turning point of the cosmic expansion
(where
$\Psi$ describes geometries on Schwarzschild type simultaneities which
contain spatial regions very close to black hole horizons) -- thus
undermining the concept of a universe evolving in time beyond its maximum
extension while containing regions of high density which would classically
very soon develop horizons and singularities. 

Similar consequences would arise from {\it all}
time-less low-entropy conditions at singularities, such as those excluding a
singular Weyl-tensor. Other conditions to eliminate
black hole singularities have recently been suggested within the framework of
loop quantum gravity~\cite{AB}. However, a quasi-classical
spacetime (that is, a quasi-trajectory of spatial geometries) would have to
be represented by an individual Everett {\it branch} of the wave function (or
by a superposition of branches which differ only by their quasi-classical
matter variables). Therefore, these quasi-classical histories of geometry
cannot {\it separately} obey unitary dynamics. In quantum gravity we are not
allowed any more to ask: What {\it happens} at the horizon? but only: What
is the structure of the wave function in the corresponding region of
configuration space?

\end{document}